\documentclass[11pt,twoside,reqno]{atmp}

\usepackage{amsmath,amssymb,amsfonts,amsthm}
\usepackage{bm,bbm}
\usepackage{graphicx}
\usepackage[T1]{fontenc}
\usepackage{esint}
\usepackage{color}
\usepackage[linktocpage,breaklinks]{hyperref}

\newcommand{\be}{\begin{equation}}
\newcommand{\ee}{\end{equation}}
\newcommand{\ba}{\begin{eqnarray}}
\newcommand{\ea}{\end{eqnarray}}
\def\bs{\begin{subequations}}
\def\es{\end{subequations}}
\def\a{\alpha}

\def\de{\delta}
\def\g{\gamma}
\def\la{\lambda}

\def\e{\epsilon}

\def\om{\omega}

\def\vr{\varrho}
\def\vp{\varphi}

\def\cA{\mathcal{A}}

\def\cD{\mathcal{D}}
\def\cE{\mathcal{E}}
\def\cF{\mathcal{F}}

\def\cI{\mathcal{I}}
\def\cJ{\mathcal{J}}
\def\cK{\mathcal{K}}
\def\cL{\mathcal{L}}

\def\cV{\mathcal{V}}
\def\bE{\mathbbm{e}}
\def\ds{d_{\rm S}}
\def\dh{d_{\rm H}}

\def\p{\partial}

\newcommand{\Eq}[1]{(\ref{#1})}
\def\com{\color{magenta}}
\def\cob{\color{blue}}
\newcommand{\oarX}[1]{\href{http://arxiv.org/abs/#1}{{\ttfamily\com #1}}}
\newcommand{\arX}[1]{\href{http://arxiv.org/abs/#1}{{\ttfamily\com arXiv:#1}}}
\newcommand{\doin}[5]{\href{http://dx.doi.org/#1}{\cob #2 {\bf #3} (#5), #4}}
\newcommand{\doij}[5]{\href{http://dx.doi.org/#1}{\cob #2 {\bf #3} (#5), #4}}
\newcommand{\ndoin}[5]{\href{#1}{\cob #2 {\bf #3} (#5), #4}}
\newcommand{\tia}[1]{\textit{#1}, }
\newcommand{\boxd}[1]{\boxed{\phantom{\Biggl(}#1\phantom{\Biggl)}}}

\def\rme{\text{e}}
\def\rmi{\text{i}}
\def\rmd{d}

\numberwithin{equation}{section}

\begin{document}

\begin{center}
\href{http://intlpress.com/site/pub/pages/journals/items/atmp/content/vols/0016/0004/00026468/index.html}{Adv.\ Theor.\ Math.\ Phys.\ 16 (2012) 1315-1348 [\arX{1202.5383}]}\\ \medskip

February 22, 2012 \hspace{2cm} AEI-2012-010
\end{center}

\barefootnote{$^*$ Present address: Instituto de Estructura de la Materia, CSIC, Serrano 121, 28006 Madrid, Spain}

\title[Momentum transforms in fractional spaces]{Momentum transforms and Laplacians in fractional spaces}

\author[Gianluca Calcagni and Giuseppe Nardelli]{Gianluca Calcagni$^{1,*}$ and Giuseppe Nardelli$^{2,3}$}
\address{$^1$ Max Planck Institute for Gravitational Physics (Albert Einstein Institute),
Am M\"uhlenberg 1, D-14476 Golm, Germany}
\addressemail{calcagni@iem.cfmac.csic.es} 
\address{$^2$ Dipartimento di Matematica e Fisica, Universit\`a Cattolica,
via Musei 41, 25121 Brescia, Italy}
\address{$^3$ INFN Gruppo Collegato di Trento, Universit\`a di Trento,
38100 Povo (Trento), Italy}
\addressemail{nardelli@dmf.unicatt.it}


\begin{abstract}
We define an infinite class of unitary transformations between position and momentum fractional spaces, thus generalizing the Fourier transform to a special class of fractal geometries. Each transform diagonalizes a unique Laplacian operator. We also introduce a new version of fractional spaces, where coordinates and momenta span the whole real line. In one topological dimension, these results are extended to more general measures.
\end{abstract}



\maketitle


\section{Introduction}

The spectral theory in fractal geometry has, by now, achieved a certain degree
of sophistication \cite{Str99,Kig01,Str06}. Given a self-similar fractal set $\cF$, one can construct a
natural Laplacian operator thereon and study its spectrum, which depends both on
the geometry (i.e., symmetries) and on the topology of the set. An open question,
however, is how to construct a  ``momentum space'' $\cF_k$ or, in other
words, whether there exists an invertible transform $F:\,\cF\to \cF_k$
generalizing the Fourier transform in $\mathbb{R}^D$. Results in this direction
were found for fractafolds \cite{StT} and post-critically finite fractals such
as the Sierpi\'nski gasket \cite{St03a,OkS}. The geometry of momentum space is,
in general, different from that of $\cF$: while $\cF$ is characterized by the
Hausdorff dimension $\dh$, some evidence is in favour of identifying, for
several fractals, the dimension of $\cF_k$ with the spectral dimension $\ds$ of
the set \cite{Akk2}. Checking the conjecture 
\be\label{conj}
\dh(\cF_k)\ \stackrel{?}{=}\ds(\cF)
\ee
is tightly related to the possibility of writing the transform $F$ explicitly.

One goal of this paper is to answer this question in the context of fractional
spaces \cite{fra4,frc1,frc2,ACOS}. Fractional spaces are continua embedded in a $D$-dimensional
manifold where ordinary calculus is replaced by fractional calculus of
fixed order. Giving up ordinary differentiability in this way guarantees that
the geometric and harmonic properties of fractional spaces have genuine fractal
features, such as anomalous dimensionality (non-integer Hausdorff and spectral
dimension) and discrete symmetries (logarithmically oscillating measures).
Allowing the fractional order to change with the scale, one obtains
multi-fractional settings endowed with a multi-fractal geometry. Here, we
construct a class of unitary transforms between position and momentum
fractional Euclidean space. If these transforms are imposed to be automorphisms, then
\be\label{dhdh} 
\dh(\cE_{\a,k}^D)=\dh(\cE_\a^D)\,.
\ee

For fractional spaces such that $\ds(\cE_\a^D)=\dh(\cE_\a^D)$ \cite{frc1}, combining with \Eq{dhdh} one would verify equation \Eq{conj} for $\cF=\cE_\a^D$. If diffusion is anomalous, however, or if momentum and position spaces are taken with different measures, $\ds\neq\dh$ and the conjecture is violated.

In multi-fractals, spectral and Hausdorff dimensions change with the probed
scale. Multi-fractional spaces realize this feature and were proposed as the
fundamental building block of field theories with improved ultraviolet (UV)
properties \cite{fra1,fra2,fra3}. A reduction of dimensionality with the physical scale has been
recognized as an agent favouring UV finiteness in the context of
quantum gravity \cite{fra1,Car09,Car10}. Dimensional flow (especially towards a
two-dimensional effective spacetime in the UV) seems to be a universal property
of independent quantum gravity models such as causal dynamical triangulations
\cite{AJL4,BeH}, asymptotically safe gravity \cite{LaR5}, spin foam dynamics
\cite{Mod08,CaM,MPM}, and Ho\v{r}ava--Lifshitz gravity \cite{Hor3,SVW1} (see also \cite{MoN}). A non-trivial fixed point with a reduced Hausdorff dimension,
associated with an anomalous scaling dimension of the metric, was recognized as
a requisite for a perturbatively renormalizable quantum gravity theory
\cite{CrS1,CrS2}. The mathematical framework of loop quantum gravity had been
developed also with the hope of realizing a fractal dimensional reduction,
before UV finiteness was rather ascribed to discreteness of the geometry.

In quantum mechanics and quantum  field theory, a
well-defined momentum space constitutes a very powerful tool for the physical
interpretation and for calculational purposes. The same holds also for fields
living in multi-fractional geometry \cite{fra4,frc2}. Before initiating a systematic construction
of a fractional field theory, it is therefore important to show the existence of a momentum
transform expanded in a basis of functions $K$ which either diagonalize the quadratic form 
$\cD K \cD K$, with $\cD$ a differential operator, or are eigenfunctions of the Laplacian operator $\cK$. If $\cD$ is self-adjoint, $\cK=\cD^2$ and the two conditions are equivalent. 
 
Since momentum transforms are specified by integral kernels $K$ that are   bounded functions, it is natural to define the momentum transform $F$ for absolutely integrable functions, $f\in L^1(X,\vr)$, for some domain $X\subset \mathbb{R}^D$ equipped with a given measure $\vr$, just as for the usual Fourier transform. In this case, $F$ is a continuous linear application from $L^1(X,\vr)$ to some subset of $C_0(\mathbb{R}^D)$, the set of continuous functions in $\mathbb{R}^D$. The more physically  interesting case of functions belonging to  $ L^2(X,\vr)$ cannot be done straightforwardly, as there are many $ L^2(X,\vr)$ functions that do not belong to  $ L^1(X, \vr)$ and for which the local definition of the integral transform does not hold. It will be possible to extend the momentum transform from $ L^1(X, \vr)\cap  L^2(X, \vr)$ onto $ L^2(X, \vr)$ only as a limiting procedure, just like for the Fourier transform. In this case, the momentum transform $F$ can be unambiguously defined with the following properties: 
\begin{enumerate}
\item[(i)] $F$ is a unitary integral transform  of $ L^2(X, \vr)$  onto itself.
\item[(ii)] $F$ can be expressed as an integral operator whose kernel is specified by the eigenfunctions of a given Laplace operator $\cK$.
\end{enumerate}
{ With the above conditions, we shall find integral representations of the Dirac distribution in terms of the eigenfunctions of $\mathcal{K}$}. Our findings will clarify the interrelation between momentum transforms and Laplacians in fractional spaces. For a given space, the momentum transform is not unique and there exist inequivalent $F$'s satisfying (i). If a particular Laplacian $\cK$ is chosen, condition (ii) can fix the momentum transform on the space, but we will end up with an infinite class of transforms and Laplacians. If we further require that
\begin{enumerate}
\item[(iii)] the Laplacian $\cK$ can be written as the square of a self-adjoint differential operator $\cD$, $\cK=\cD^2$,
\end{enumerate}
then both the momentum transform and $\cK$ are uniquely defined for a given fractional space.

Section \ref{feft} briefly introduces multi-fractional spaces, both in their original formulation (``unilateral'') and in a novel ``bilateral'' version. In the former (Section \ref{mfuni}), the measure has support over the first orthant of $\mathbb{R}^D$. Both coordinates $x$ and momenta $k$ will be  non-negative. In the bilateral case (Section \ref{mfbil}), the measure weight is a function of the absolute value of the coordinates, and the support of the measure is the whole space: coordinates and momenta can take both signs. For each version, we define two inequivalent second-order Laplacians which, respectively, have played and will play a major role in the formulation of the theory.

An infinite class of momentum transforms in the unilateral and bilateral versions is constructed in Section \ref{fmbt}. The class is parametrized by a parameter $l$, which is continuous in the unilateral case but can take only discrete values in the bilateral world. The requirement (iii) fixes $l$ once and for all to a special value. The multi-fractional and complex fractional cases are also discussed. Section \ref{conc} is devoted to conclusions.


\section{Multi-fractional Euclidean spaces}\label{feft}


\subsection{Unilateral world}\label{mfuni}

Let $\mathbb{R}^D_+$ be the first orthant of Euclidean space in $D$ (integer)
topological dimensions. Define the fractional measure
\bs\label{mea}\be\label{frames}
\rmd\vr_\a(x)=\rmd^Dx\,v_\a(x)\,,
\ee
where the ``isotropic'' measure weight is
\be\label{frames2}
v_\a(x)=\prod_{\mu=1}^D v_\a(x^\mu):=\prod_{\mu=1}^D \frac{(x^\mu)^{\a-1}}{\Gamma(\a)}\,,
\ee\es
$x^\mu\geq 0$ are $D$ coordinates, $\Gamma$ is the gamma function, and $1/2\leq\a\leq 1$ is a real parameter. The measure is isotropic in the fractional charge $\a$, but anisotropic measures with different $\a_\mu$ are also possible. We do not consider the anisotropic case for simplicity and also because isotropic fractional spaces are sufficient to realize the physics outlined in \cite{fra4,frc2}.

The volume of a $D$-ball of radius $R$ scales as
\be
\cV^{(D)}=\int_{D{\rm -ball}}\rmd\vr_\a(x)\propto R^{\dh}\,,\qquad \dh=D\a\,,
\ee
where $\dh$ is the Hausdorff dimension of the space $\cE_\a^D$ endowed with the
measure \Eq{mea}. Summing or integrating over all possible values of $\a$, weighted
by a factor $g_\a$, one obtains the multi-fractional measure
\be\label{mume}
\rmd\vr(x)=\sum_\a g_\a\rmd\vr_\a(x)\,,
\ee
representing a space $\cE_*^D$ whose dimension changes with the scale. In fact,
the sum or integral in $\a=\a(\ell)$ can be regarded as over a scale $\ell$
increasing with $\a$ \cite{frc2}.

Given a Lagrangian density $\cL$, which may or may not depend on $\a$, the
fractional action reads
\be\label{act0}
S=\sum_\a g_\a S_\a\,,\qquad S_\a = \int_0^{+\infty}\rmd\vr_{\a}(x)\, \cL\,.
\ee
For a real scalar field, $\cL=\phi \cK \phi/2-V(\phi)$, where $\cK$ is a kinetic operator, $V$ is a potential and $\phi$ has scaling dimension $[\phi]=(\dh-[\cK])/2$. This vanishes at the critical point $\a_*=[\cK]/D$, signalling power-counting renormalizability \cite{frc2,fra1,fra2}. If we identify $\a_*$ with the lowest value $\a=1/2$ in a theory with Lorentzian signature, at the lowest scale in the dimensional flow spacetime has dimension ${\dh}_*=[\cK]=D/2$, and if $D=4$ then ${\dh}_*=2$.

In general, the Hausdorff and spectral dimension in the UV depend on the choice of Laplacian. This choice determines uniquely the invertible unitary transform (possibly parametrized as a class of transforms) linking fractional position and momentum spaces. A two-dimensional UV limit for effective spacetimes is typical in quantum gravity models, so we are mainly interested in second-order differential operators $\cK$. In \cite{frc1,frc2}, the following Laplacian was used:
\be\label{ck2}
\cK_1:=\frac{1}{v_\a}\de^{\mu\nu}\p_\mu(v_\a\p_\nu\,\cdot\,) =
\de^{\mu\nu}\left(\p_\mu\p_\nu-\frac{1-\a}{x^\mu}\,\p_\nu\right)\,,
\ee
where Einstein's summation convention is assumed and $\de^{\mu\nu}$ is the Kronecker delta. In spaces with Lorentzian signature, this is replaced by the Minkowski metric $\eta^{\mu\nu}$. The analogy between the measure weight $v_\a$ and the determinant of the metric $\sqrt{g}$ in a Riemannian space makes equation \Eq{ck2} resemble the covariant Laplacian. Another possibility, which we introduce here, is to consider the operator
\bs\label{cknew}\ba
\cK_2 &:=& \frac{1}{\sqrt{v_\a}}\de^{\mu\nu}\p_\mu\p_\nu(\sqrt{v_\a}\,\cdot\,)\\
 &=&\de^{\mu\nu}\left[\p_\mu\p_\nu-\frac{1-\a}{x^\mu}\,\p_\nu+\frac{(1-\a)(3-\a)}{4}\frac{1}{x^\mu x^\nu}\right]\,.
\ea\es
Notice the extra centrifugal potential term. In the limit $\a\to 1$, $\cK_1=\cK_2=\p^2$.


\subsection{Bilateral world}\label{mfbil}

The action \Eq{act0} is defined with a measure whose support is the positive
orthant of $\mathbb{R}^D$ \cite{fra4,frc1,frc2,ACOS}. The choice $x^\mu\geq 0$ is made so that this
measure is real-valued and does not pick
complex phases arising when one changes orthant for some $x^\mu\to -x^\mu$. 
 Consider the integral
\be\nonumber
\int_0^{+\infty}\rmd^D x\, v_\a(x)\, f(x)\,,
\ee
where $f$ is any function such that the integral is well-defined. Splitting the
integral artificially in two and changing variable $x\to -x$ in the second piece, one
finds that
\be\label{modu}
\int_0^{+\infty}\rmd^D x\, v_\a(x)\, f(x) = \int_{-\infty}^{+\infty}\rmd^D x\,
v_\a(|x|)\, \frac{f(|x|)}{2}\,,
\ee
where
\be\label{memu}
v_\a(|x|)=\prod_{\mu=1}^D v_\a(|x^\mu|):=\prod_{\mu=1}^D \frac{|x^\mu|^{\a-1}}{\Gamma(\a)}\,.
\ee
Equation \Eq{modu} states that unilateral fractional integrals defined on a
functional space of arbitrary (but ``good'') functions are equivalent to
bilateral fractional integrals defined on a functional space of even functions.
Conversely, a bilateral world defined on a functional space with indefinite
parity is equivalent to an unilateral one with a functional space of even
functions.

At first sight, there seems to be little point in an exercise stating a simple
mathematical equivalence. Using one or the other formulation should be just a
matter of convention. Unilateral fractional measures might seem preferable over those with the absolute value,
considered in \cite{Tar2} and \cite[Section 2.5]{frc1}, since the latter have the
small disadvantage that they hide the integrable singularity at $x^\mu=0$.
However, a careful inspection of the physics one can do in fractional spaces
shows that unilateral worlds (including the $\a=1$ case) may be problematic for a
sensible formulation of quantum mechanics \cite{CNS}. Therefore, the bilateral version of
fractional spaces is equation \Eq{memu} with integration support $\mathbb{R}^D$, 
\be\label{unibi}
\int_0^{+\infty}\rmd^D x\, v_\a(x)\to \int_{-\infty}^{+\infty}\rmd^D x\,
v_\a(|x|)\,,
\ee
which must be considered as a quite distinct implementation of fractional geometry. For simplicity, we shall
use the same symbol $v_\a(x)$ for the measure weight \Eq{memu}, as the difference is explicit
in the integration range. In fact, $x^\mu\geq 0$ in the unilateral world and one could have taken directly equation \Eq{memu}, the absolute value being pleonastic in this case. Then, the definitions \Eq{ck2} and \Eq{cknew} are unaltered.


\section{Fractional momentum transforms}\label{fmbt}

In this section, we show that:
\begin{enumerate}
\item[(1)] In the unilateral case, there exists an infinite number of invertible unitary transforms $F_\a^l$ parametrized by a parameter $l$ such that ${\rm Re}(l)>-1$. Each transform is 
realized by an integral kernel formed by the
 eigenfunctions $c_\a^l$ of a second-order Laplacian operator $\cK_{\a,l}$.
\item[(2)] Only the transform with $l=1/2$ is such that the associated Laplacian can be written as $\cK=\cD^2$, where $\cD$ is a first-order self-adjoint differential operator.
\item[(3)] In the bilateral case, there exists an infinite number of invertible unitary transforms such that $l=n-1/2$ is half-integer.
\end{enumerate}


\subsection{General setting}\label{gese}

Without loss of generality, in this subsection we shall consider the $D=1$ dimensional case; the extension to $D$ dimensions will be straightforward.  Let $F$ be a prototype of momentum (linear) transformation specified by a bounded kernel $K$. Acting on a function $f(x)$, the $F$ transform is
\be
\label{mt1}
\tilde f (k) = \int \rmd \vr(x)\, K(k,x) f(x) =: F[f(x)]\,.
\ee
If $K$ is  bounded, it does not present singularities along the integration path and  asymptotically tends to  some constant.\footnote{The case of power-like bounded kernels can be treated along the same lines, either modifying the function space or modifying the measure $\vr(x)$.} Then, equation \Eq{mt1} is well defined for any function 
$f(x)\in L^1(\mathbb{R}, \vr)$. The image of the transform \Eq{mt1} is a not-so-easily characterizable subset of $C_0(\mathbb{R})$, the set of the continuous functions on $\mathbb{R}$. Thus, $F$ is a linear operator from  $L^1(\mathbb{R}, \vr)$ into $C_0(\mathbb{R})$.

In the context of quantum mechanics, equation \Eq{mt1} meets with two problems: (a) it should be defined in $L^2(\mathbb{R}, \vr)$, rather than in $ L^1(\mathbb{R}, \vr)$; (b) its image should be $ L^2(\mathbb{R}, \vr)$, and not a subset of $C_0(\mathbb{R})$. To proceed, one should first notice that, for any $f_\kappa(x)$ continuous and defined on a compact support $\kappa$, $F[f_\kappa(x)]\in L^2(\mathbb{R}, \vr)$. Since $L^2(\mathbb{R}, \vr)$ is closed, the idea is to define the $F$ transform for a generic function in $L^2(\mathbb{R}, \vr)$ as the limit, in the  $L^2(\mathbb{R}, \vr)$ topology, of a sequence of functions defined on a compact support. For instance, to any continuous  $f(x)\in L^2(\mathbb{R}, \vr)$ we could associate the following sequence $f_n\to f$:
\be
\label{mt2}
f_n(x) = \begin{cases} f(x)  \ \ &\text{if} \ |x|\le n\,,\\
0\ \ &\text{if} \ |x|>n\,, \end{cases}
\ee
and define the $F$ transform of $f$ as the limit of the $F$-transformed sequence
\be
\label{mt3}
\tilde f(k) := \lim_{n\to \infty} \tilde f_n(k) = \lim_{n\to \infty} \int_{-n}^{+n} d\vr(x)\, K(k,x) f(x)\,.
\ee
Since the space of the continuous functions on $\mathbb{R}$ is dense  in $L^2(\mathbb{R}, \vr)$,  the limit procedure defined in equation \Eq{mt3} provides  a map from $L^2(\mathbb{R}, \vr)$ to $L^2(\mathbb{R}, \vr)$. The limit is understood in the $L^2$ topology.  Consequently,  if $f(x)\in L^1(\mathbb{R}, \vr)\cap L^2(\mathbb{R}, \vr)$, $\tilde f(k)$ defined as in \Eq{mt1} could be different from that obtained through \Eq{mt3}. However, they belong to the same equivalence class ({i.e.}, they are equal almost everywhere). From now on, we shall interpret all the equalities in the $L^2$ sense, understanding the above limiting procedure, if needed. Also, for any $f$ and $g$ in $L^2(\mathbb{R},\vr)$ we define the inner product
\be
(f,\,g):=\int\rmd\vr(x)\, f^*(x)\,g(x)\,,\qquad f,g\in L^2_\a\,,
\ee
where $*$ denotes complex conjugation. The norm of a functions is then $\|f\|^2:=(f,\,f)$.

If the map is invertible, there should exist one $K^{-1}(k,x)$ such that
\be
\label{mt4}
f (x) = \int \rmd\tau (k)\, K^{-1}(k,x) \tilde f(k) =: F^{-1}[\tilde f(k)]\,,
\ee
where the integration measure $\tau(k)$ in momentum space is allowed to be different from $\vr(x)$. Imposing $F[F^{-1}[\tilde f(k)] ]= \tilde f(k)$, one obtains the resolution of the identity in terms of $K$:
\be
\label{mt5}
\int \rmd\vr (x)\, K(k,x) K^{-1} (k',x)  = \delta_\tau (k,k')\,,
\ee
where the distribution $\delta_\tau$ is the delta distribution when the momentum measure is $\tau$, {i.e.},
\be
\label{mt6}
\int \rmd\tau(k')\, f(k') \delta_\tau (k,k') = f(k)\,.
\ee 
Since it must also be $F^{-1}[F[f(x)]]= f(x)$, one obtains, in position space, a different representation of the
delta distribution,
\be
\label{mt7}
\int \rmd\tau(k)\, K(k,x) K^{-1} (k,x')  = \delta_\vr(x,x')\,,
\ee
such that
\be
\label{mt7b}
\int \rmd\vr(x')\, f(x') \delta_\vr (x,x') = f(x)\,.
\ee 
At this point, if we require that the momentum transform be an automorphism, then the partition of the unity is unique (the two representations of the delta are equal), yielding $\tau=\vr$. The general case will be commented on in the conclusions. Then, both the kernels $K$ and $K^{-1}$ must depend on the product $kx$; in particular, they are invariant under the exchange $(k,x)\leftrightarrow (x,k)$ and the identities \Eq{mt5} and \Eq{mt7} are equivalent with $k$ and $x$ switched.

In ordinary  quantum mechanics, coordinate and momentum representations are equivalent pictures for describing a physical system: the Fourier transform not only maps any $L^2$ element into another $L^2$ element, but it is also a surjective map, that is, any $L^2$ element can be seen as a Fourier transform of another $L^2$ element. This is guaranteed  by the fact that the inverse Fourier transform is itself a Fourier transform. Precisely the  same will happen in our case and  the  momentum transforms $F$ we shall define  are ``onto'' $L^2(\mathbb{R}, \vr)$. Consequently, unitarity of the $F$ transform solely depends on the Parseval identity. If the latter holds,
\be
\label{mt8}
\|\tilde f(k) \| = \|f(x)\|\,,
\ee
then the transformation is unitary. In fact,
\be\nonumber
(f,f) = (\tilde f , \tilde f) = (Ff,Ff) = (F^\dagger F f, f)
\ee
implies
\be
F^\dagger = F^{-1} \qquad \text{or} \qquad K^*(kx)= K^{-1}(kx)\,,
\ee
and the momentum transform is unitary. The converse is also true: by reading the above passages in the opposite direction, if $F$ is unitary then equation \Eq{mt8} is satisfied. Thus, unitarity in the transformations we are going to present can be always verified by checking the validity of the Parseval identity.


\subsection{Fourier transform}

Consider Euclidean space $\mathbb{R}^D$, where $D$ is the topological (integer) dimension. The
direct and inverse Fourier transforms of a function $f(x)\in L^2(\mathbb{R})$ are
\bs\label{ft}\ba
\tilde f(k) &:=& \frac{1}{(2\pi)^{\frac{D}{2}}}\int_{-\infty}^{+\infty}\rmd^Dx\,
f(x)\,\rme^{-\rmi k\cdot x}=:F_1[f(x)]\,,\\
f(x) &=& \frac{1}{(2\pi)^{\frac{D}{2}}}\int_{-\infty}^{+\infty}\rmd^Dk\, \tilde f(k)\,\rme^{\rmi k\cdot x}=:F^{-1}_1[\tilde f(k)]\,,
\ea\es
where $ k\cdot x=k_\mu x^\mu=k_1 x_1+\dots+k_D x_D$. $F$ is invertible and unitary, and the Dirac distribution admits an integral representation in terms of the Fourier kernel $K(kx) = \rme^{\rmi k\cdot x}$:
\be\label{die}
\frac{1}{(2\pi)^D} \int_{-\infty}^{+\infty}\rmd^D k\,\rme^{\pm \rmi k\cdot (x-x')}=\delta(x-x')\,.
\ee
Notice that the $D$-dimensional kernel $K$ is just the product of $D$ factorized kernels $K_{(\mu)} = \rme^{\rmi k_\mu x_\mu}$ ($\mu$ not summed), one for each dimension $\mu = 1,\ldots ,D$.
It is thanks to equation \Eq{die} that Parseval relation \Eq{mt8} holds, and the Fourier transform is unitary.
From equation \Eq{ft}, $F_1^{-1}[\tilde f(k)] = F_1^{}[\tilde f(-k)]$. Consequently, any $f(x)\in L^2(\mathbb{R})$ is the Fourier transform of $g(x) = F_1[f(-k)] \in  L^2(\mathbb{R})$.

The Fourier transform in $\mathbb{R}^D_+$ is expanded in cosines or sines rather than phases. From
equations \Eq{ft} and \Eq{die},
\ba
1 &=&\int_{-\infty}^{+\infty}\rmd^D x\, \delta(x)\,\rme^{-\rmi k\cdot x}\nonumber\\
&=&(2\pi)^\frac{D}{2}\int_{0}^{+\infty}\rmd^D x\, \delta(x)\,c(k,x)\,,\label{dirc1}
\ea
where
\ba
c(k,x)&:=&\left(\frac2\pi\right)^\frac{D}{2}\prod_\mu\cos(k_\mu
x_\mu)\nonumber\\
&=&\left(\frac2\pi\right)^\frac{D}{2}\cos(k_1 x_1)\dots\cos(k_D x_D)\,.\label{cokx}
\ea
Similarly,
\be\label{dirc2}
\delta(x-x') = \frac{1}{(2\pi)^\frac{D}{2}}\int_{0}^{+\infty}\rmd^D k\, c(k,x-x')\,.
\ee
Also the inverse transform runs over positive values of the integration
variable. Equations \Eq{dirc1}--\Eq{dirc2} completely define the transformation
properties of the delta distribution in unilateral representation. Then, for a function $f$,
\bs\label{ctr}\ba
\tilde f(k) &=& \int_{0}^{+\infty}\rmd^D x\, f(x)\,c(k,x)=:
F_c[f(x)]\,,\label{ctr1}\\
f(x) &=& \int_{0}^{+\infty}\rmd^D k\, \tilde f(k)\,c(k,x)= F^{-1}_c[\tilde f(k)]\,.\label{ctr2}
\ea\es
Plugging the first equation into the second, one notices that (for each
direction) $2\cos(kx) \cos(kx')=\cos[k(x'-x)]+\cos[k(x'+x)]$, and performs the
integration via equation \Eq{dirc2}. In $D$ dimensions, this gives
\be\nonumber
f(x) = \int_{0}^{+\infty}\rmd^D x'\,[\de(x'-x)+\de(x'+x)] f(x')\,.
\ee
The support of the second delta is outside the integration range for any $x'>0$,
and that contribution vanishes. Therefore, we have the following resolutions of the identity:
\bs\label{ccdelta}\ba
\delta(x-x') &=& \int_{0}^{+\infty}\rmd^D k\, c(k,x)\,c(k,x')\,,\\
\delta(k-k') &=& \int_{0}^{+\infty}\rmd^D x\, c(k,x)\,c(k',x)\,,
\ea\es
where the second equation comes from the first under the exchange $k\leftrightarrow x$.
Equations \Eq{ccdelta} guarantee the validity of the Parseval identity and then  unitarity of the cosine Fourier transform follows. Notice that the inverse of the cosine Fourier transform is itself, $F_c^{-1} = F_c$ and surjectivity follows.

The sine transform
\bs\label{sft}\ba
\tilde f(k) &=& \int_{0}^{+\infty}\rmd^D x\, f(x)\,s(k,x)=:F_s[f(x)]\,,\\
f(x) &=& \int_{0}^{+\infty}\rmd^D k\, \tilde f(k)\,s(k,x)\,,\label{sft2}
\ea\es
where
\ba
s(k,x)&:=&\left(\frac2\pi\right)^\frac{D}{2}\prod_\mu\sin(k_\mu
x_\mu)\nonumber\\
&=&\left(\frac2\pi\right)^\frac{D}{2}\sin(k_1 x_1)\dots\sin(k_D x_D)\,,\label{sokx}
\ea
can be developed along the same lines. Upon repeating the above inversion
argument with the cosine functions replaced by the sines, one ends up with $2\sin(kx) \sin(kx')=\cos[k(x'-x)]-\cos[k(x'+x)]$
that,  using again  equation \Eq{dirc2}, leads to the resolutions of the identities  
\bs\label{ssdelta}\ba
\delta(x-x') &=& \int_{0}^{+\infty}\rmd^D k\, s(k,x)\,s(k,x')\,,\\
\delta(k-k') &=& \int_{0}^{+\infty}\rmd^D x\, s(k,x)\,s(k',x)\,,
\ea\es
and unitarity immediately follows.

The choice between cosine and sine transform typically depends on the behaviour
of the functions $f(x)$ at the origin. If $f(0)=0$, the sine expansion is chosen
for the sole purpose of taking equation \Eq{sft2} at face value, i.e., as a pointwise
equality (then, expanding around $x\sim 0$ one does not meet with
contradictions). However, this is not strictly necessary, as the equalities in   \Eq{ctr} and \Eq{sft}
are intended globally, in the $L^{2}(\mathbb{R}^D_+)$ norm and, as such, they correspond to pointwise equalities only almost everywhere.\footnote{Sometimes, the cosine and sine transforms are presented as the ``natural'' transform for,
respectively, even and odd functions. We have just seen that they work perfectly
well for general functions, not only those with definite parity. More precisely,
in a unilateral world there is no notion of parity, and the correct statement
is that cosine/sine transforms are well defined also for functions with definite
parity when analytically continued to the negative semi-axis.}


\subsection{Unilateral world}\label{btuni}

\subsubsection{Fractional Bessel transforms}\label{frabe}

To obtain a partition of the identity in the fractional case, we must be able to express the (fractional analog of the) Dirac distribution as an integral representation in terms of the kernel of the transform. The fact that fractional coordinates are non-negative suggests the following strategy, which is not the most natural but it will define the correct eigenfunctions of the operator \Eq{ck2}. In $n$-dimensional Euclidean space, the radial delta distribution $\de(r)$ carries a scaling dimension
$[\de(r)]=-1$, where
\be\nonumber
r:=\sqrt{ x_1^2+\dots+x_n^2}\,.
\ee
For radial functions $f(r)$, the $n$-dimensional Fourier transform in
hyperspherical coordinates reduces to a one-dimensional, $n$-dependent
invertible transform in $r$. Analytic continuation of this transform to
non-integer values $n\to\a$ leads to the transform in a fractional space with
$\dh=\a$, where we find a distribution $\de_\a$ such that $[\de_\a]=-\a$.
Repeating the argument for all  the $D$ directions yields the final result.

Taking the Fourier transform \Eq{ft} in $D\to n$ dimensions, we move to hyperspherical
coordinates,
\be\nonumber
(x_1,\cdots,x_n)\to(r,\vp,\theta_1,\cdots\theta_{n-2})\,,
\ee
so that the coordinate transformation is
\ba
&& x_1=r\sin\vp \prod_{j=1}^{n-2}\sin\theta_i\,,\qquad x_2=r\cos\vp
\prod_{j=1}^{n-2}\sin\theta_j\,,\nonumber\\
&& x_{j+2}=r\cos\theta_j \prod_{l=j+1}^{n-2}\sin\theta_l\,,\qquad
x_n=r\cos\theta_{n-2}\,,\nonumber
\ea
and the measure reads
\be\nonumber
\rmd^n x= \rmd r\,r^{n-1}\,\rmd\vp\,\prod_{j=1}^{n-2}\rmd\theta_j\,(\sin\theta_j)^i\,.
\ee
Choose the orientation of the  ${\bf x}$ frame such that the (fixed) vector  ${\bf k}$ has components $(0,\cdots,0,k_n)$, so that ${\bf k}\cdot {\bf x}=kx_n=kr\cos\theta_{n-2}$, where $k:=|{\bf k}|$. Suppose $f$ is a function only of $r$. Then the Fourier transform $\tilde f$ is a function only of $k$ and
\ba
\tilde f(k) &=& \frac{1}{(2\pi)^{\frac{n}{2}}}\int_{0}^{+\infty}\rmd
r\,r^{n-1}\int_0^{2\pi}\rmd\vp\prod_{j=1}^{n-3}\int_0^\pi\rmd\theta_j\,
(\sin\theta_j)^j\nonumber\\
&&\times\int_0^\pi\rmd\theta_{n-2}\,(\sin\theta_{n-2})^{n-2}f(r)\,\rme^{-\rmi
kr\cos\theta_{n-2}}.\label{raf}
\ea
From formul\ae\ 3.621.1, 8.335.1 and 8.384.1 of \cite{GR},
\be\nonumber
\prod_{j=1}^{n-3}\int_0^\pi\rmd\theta_j\,(\sin\theta_j)^j=\prod_{j=1}^{n-3}\frac
{\sqrt{\pi}\,\Gamma\left(\frac{j+1}{2}\right)}{\Gamma\left(\frac{j+2}{2}\right)}
=\frac{\pi^{\frac{n-3}{2}}}{\Gamma\left(\frac{n-1}{2}\right)}\,,
\ee
while \cite[formula 8.411.7]{GR}
\be
\int_0^\pi\rmd\theta_{n-2}\,(\sin\theta_{n-2})^{n-2}\,\rme^{-\rmi
kr\cos\theta_{n-2}}=\sqrt{\pi}\,2^{\frac{n}{2}-1}\frac{\Gamma\left(\tfrac{n-1}{2
}\right)}{\Gamma(n)}\,c_n(kr)\,,\nonumber
\ee
where
\be
c_n(kr):=\Gamma(n)(kr)^{1-\frac{n}{2}}J_{\frac{n}{2}-1}(kr)\,.
\ee
Here, $J_\nu$ is the Bessel function of the first kind:
\be\label{Jnu}
J_\nu(z)= z^\nu \cJ_\nu(z):=z^\nu\sum_{m=0}^{+\infty} \frac{(-1)^m}{2^{2m+\nu}
m!\Gamma(m+\nu+1)}\,z^{2m}\,,
\ee
from which it follows that $c_n$ is even. Then, equation \Eq{raf} and its inverse
become
\bs\label{betr}\ba
\tilde f(k) &=& k^{1-\frac{n}{2}}\int_0^{+\infty}\rmd
r\,r^{\frac{n}{2}}\,f(r)\,J_{\frac{n}{2}-1}(kr)\,,\label{fk1}\\
       f(r) &=& r^{1-\frac{n}{2}}\int_0^{+\infty}\rmd k\,k^{\frac{n}{2}}\,\tilde
f(k)\,J_{\frac{n}{2}-1}(kr)\,,\label{fk2}
\ea\es
which can be rewritten as
\bs\label{fkc}\ba
\tilde f(k) &=& \int_0^{+\infty}\rmd
r\,\frac{r^{n-1}}{\Gamma(n)}\,f(r)\,c_n(kr)\,,\label{fkc1}\\
       f(r) &=& \int_0^{+\infty}\rmd k\,\frac{k^{n-1}}{\Gamma(n)}\,\tilde
f(k)\,c_n(kr)\,.\label{fkc2}
\ea\es
A self-consistency check is to show that $f(r)$ is the inverse of $\tilde f(k)$.
Plugging equation \Eq{fk1} into \Eq{fk2},
\ba
r^{\frac{n}{2}}f(r) &=& r\int_0^{+\infty}\rmd
k\,k^{\frac{n}{2}}J_{\frac{n}{2}-1}(kr)\,\tilde f(k)\nonumber\\
 					&=& r\int_0^{+\infty}\rmd
k\,k\,J_{\frac{n}{2}-1}(kr)\int_0^{+\infty}\rmd
r'\,{r'}^{\frac{n}{2}}\,f(r')\,J_{\frac{n}{2}-1}(kr')\nonumber\\
                      &=& \int_0^{+\infty}\rmd
r'\,\de(r-r')\,{r'}^{\frac{n}{2}}\,f(r')\,,\label{check}
\ea
where in the last line we used the integral representation of the Dirac
distribution in terms of Bessel functions \cite[equation 1.17.13]{NIST},
\be\label{denist}
\de(r-r')=r\int_0^{+\infty}\rmd
k\,k\,J_{\frac{n}{2}-1}(kr)J_{\frac{n}{2}-1}(kr')\,,\qquad \forall~n>0\,.
\ee

The generalization of the pair \Eq{fkc} to a one-dimensional fractional space is
straightforward upon the substitutions $r\rightarrow x$ and $n\rightarrow\a$.
The resulting measure in position space is correct and, because
equation \Eq{denist} is valid for any complex number $l=n/2-1$ such that ${\rm Re}(l)>-1$,
$f(x)$ is indeed the inverse transform. For this very reason, in $D$ topological dimensions there exists a whole class of fractional Bessel transforms $F_\a^{l}$ of a function $f(x)=f(x_1,\dots,x_D)$:
\bs\label{fblt}\ba
\tilde f(k) &:=& \int_0^{+\infty}\rmd\vr_\a(x)\, f(x)\,c_\a^l(k,x)=:
F_\a^l[f(x)],\label{fbl1}\\
f(x) &=& \int_0^{+\infty}\rmd\vr_\a(k)\,\tilde f(k)\,c_\a^l(k,x)\,,\label{fbl2}
\ea\es
where the basis functions are
\ba
c_\a^{l}(k,x) &:=& \prod_\mu c_{\a,\mu}^{l}(kx)\nonumber\\
&:=& \prod_\mu\sqrt{\frac{k^\mu x^\mu}{v_\a(k^\mu)v_\a(x^\mu)}}\, J_{l}(k^\mu x^\mu)\,.\label{cakx}
\ea
Equation \Eq{fbl2} corresponds to the case where momentum space has the same geometry as position space. In particular, equation \Eq{dhdh} holds. The transform \Eq{fblt} is a generalization of the  Bessel (also called Hankel)  transform (e.g., \cite{Deb}).

From equation \Eq{denist}, the integral representation of the ``fractional'' Dirac
distribution is
\bs\ba
\de_\a(x,x')&:=&\frac{\delta(x-x')}{\sqrt{v_\a(x)v_\a(x')}}\\
&=&\int_0^{+\infty}\rmd\vr_\a(k)\,
c_\a^l(k,x)c_\a^l(k,x')\,,
\ea\label{deall}\es
which has the expected scaling dimension and is not translation invariant. That
$\de_\a$ plays the role of the Dirac distribution in fractional geometry is
clear from a check identical to equation \Eq{check}, leading to
\be\label{frad}
f(x)=\int_0^{+\infty}\rmd\vr_\a(x')\,\delta_\a(x,x')f(x')\,.
\ee
Equation \Eq{deall} permits to prove the Parseval relation associated with the integral transform \Eq{fblt}, and therefore the unitarity of the whole family of transforms $F_\a^l$.
Note that the inverse transform is equal to the direct transform, and $F_\a^l$ is surjective for any $l$.

 The definition $\de_\a(x,x'):=v_\a^{-1}(x)\delta(x-x')$ was already guessed in
\cite{fra2} for a space with general Lebesgue--Stieltjes measure, but the integral representation found here is its rigorous expression in fractional spaces.

\subsubsection{Laplacians and quadratic form}\label{laqua}

The momentum transform  previously discussed is not unique and we found an infinite class $F_\a^l$. A specific choice of Laplacian operator selects a finite number of transforms. In our case, the family of kinetic operators
\bs\label{ckal}\ba
\cK_{\a,l} &:=&\sum_\mu \frac{x_\mu^{l-\frac12}}{\sqrt{v_\a(x_\mu)}}\, \p_\mu \left\{x_\mu^{1-2l} \p_\mu\left[x_\mu^{l-\frac12}\sqrt{v_\a(x_\mu)}\, \,\cdot\,\right]\right\}\label{ckasv}\\
&=&\sum_\mu x_\mu^{l-\frac{\a}{2}} \p_\mu \left[x_\mu^{1-2l} \p_\mu\left(x_\mu^{l-1+\frac{\a}{2}} \,\cdot\,\right)\right]\\
 &=&\sum_\mu\left[\p_\mu^2-\frac{1-\a}{x^\mu}\,\p_\mu+\frac{(2-\a)^2-4l^2}{4x_\mu^2}\right]\,,
\ea\es
is engineered so that the kernel of the transform $F_\a^l$ yields the two solutions of the eigenvalue equation
\bs\ba
&&\left(\cK_{\a,l}+k^2\right)c_\a^{\pm l}(k,x)=0\,,\\
&&k^2 := k_\mu k^\mu= (k_1)^2+ \dots+(k_D)^2\,.
\ea\es
In particular, the operators \Eq{ck2} and \Eq{cknew} are
\ba
\cK_1 &=& \cK_{\a,1-\frac{\a}{2}}\,,\qquad l=1-\frac{\a}{2}\,,\label{k1c}\\
\cK_2 &=& \cK_{\a,\frac{1}{2}}\,,\qquad l=\frac{1}{2}\,.\label{k2c}
\ea
The transform $F_\a^{-l}$ with $l=1-\a/2$ was employed in a companion paper \cite{frc1} to calculate the heat kernel and the spectral dimension of $\cE_\a^D$. Since the order of $\cK_{\a,l}$ is the same for any $l$, all the results of \cite{fra4,frc1,frc2} concerning the spectral dimension are unaffected by the present discussion.

\begin{figure}
\centering
\includegraphics[width=9cm]{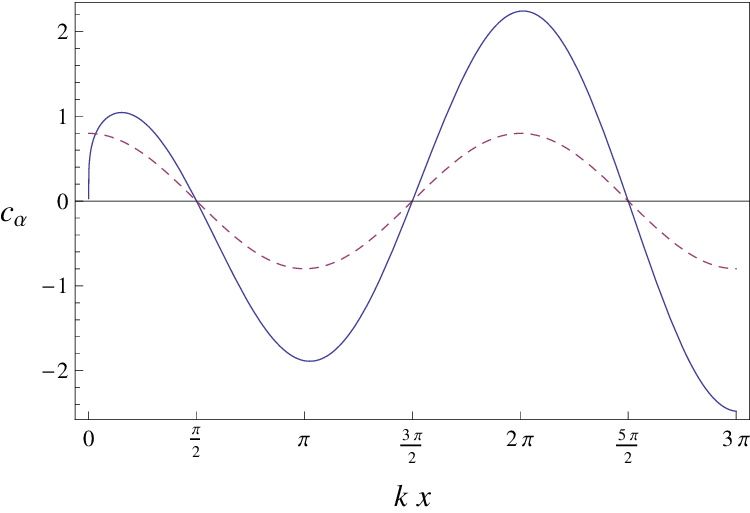} \hspace{.5cm}
\includegraphics[width=9cm]{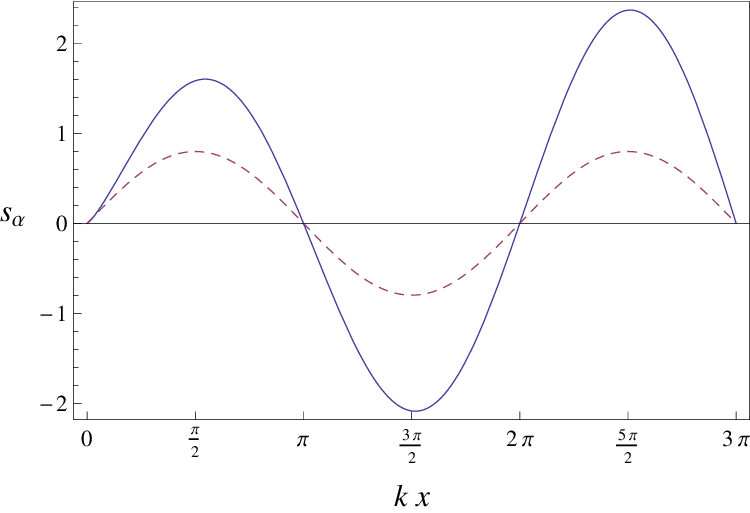}
\caption{\label{fig1}
The functions $c_\a$ (thick curve, top panel) and $s_\a$ (thick curve, bottom panel) for $\a=1/2$. The dashed curves are $c_1=c$ and $s_1=s$, respectively.}
\end{figure}

Here, however, we are interested in a more natural choice of the parameter $l$ which will allow us to write the Laplacian operator as the square of a self-adjoint derivative operator. With the same value of $l$ it is also possible to extend the transform to bilateral fractional spaces. This transform is associated with the kinetic operator \Eq{cknew}. Notice that, when $\a=1=2l$, $c_1^{-1/2}(k,x)=c(k,x)$ and $c_1^{1/2}(k,x)=s(k,x)$ (see \cite[equation (10.16.1)]{NIST} and equation \Eq{cokx}) and equations \Eq{fblt} reduce to the ordinary cosine and sine Fourier transforms \Eq{ctr} and \Eq{sft}. For general $\a$, the eigenfunctions of $\cK_2$ are
\bs\label{cakx12}\be
c_\a^{-\frac12}(k,x) =\left(\frac{2}{\pi}\right)^{\frac{D}{2}}\prod_\mu\frac{\cos(k^\mu x^\mu)}{\sqrt{v_\a(k^\mu)v_\a(x^\mu)}}=:c_\a(k,x)\,,
\ee
\be
c_\a^{\frac12}(k,x) =\left(\frac{2}{\pi}\right)^{\frac{D}{2}}\prod_\mu\frac{\sin(k^\mu x^\mu)}{\sqrt{v_\a(k^\mu)v_\a(x^\mu)}} =:s_\a(k,x)\,.
\ee\es
These functions are shown in figure \ref{fig1}. They vanish in $x=0$ and their amplitude increases as a mild power law.

The $l=1/2$ case is special because it is the only one where $\cK_{\a,l}$ is the square of a first-order differential operator. Consider the integral
\be\label{lk2}
-\int_0^{+\infty}\rmd\vr_\a(x)\,\de^{\mu\nu}\cD_{\mu}^{\a,l} f(x)\, \cD_\nu^{\a,l} f(x)\,,
\ee
where (no sum over $\mu$)
\bs\label{quaf}\ba
\cD_\mu^{\a,l} &:=& \frac{x_\mu^{\frac12-l}}{\sqrt{v_\a(x_\mu)}}\, \p_\mu \left[x_\mu^{l-\frac12}\sqrt{v_\a(x_\mu)}\,\cdot\,\right]\label{quaf0}\\
               &=& \frac{1}{x_\mu^{l-1+\frac{\a}{2}}}\, \p_\mu \left(x_\mu^{l-1+\frac{\a}{2}}\,\cdot\,\right)\,.
\ea\es
We can take the case $D=1$ for simplicity. Integrating by parts, one obtains
\be\label{ddk}
-\int_0^{+\infty}\rmd\vr_\a(x)\,\cD^{\a,l} f(x)\, \cD^{\a,l} f(x)= \int_0^{+\infty}\rmd\vr_\a(x)\,f(x)\,\cK_{\a,l}f(x)\,,
\ee
provided the following boundary term vanishes:
\be
\lim_{\e\to0^+} x^{\frac{\a}{2}-l} f(x) \p\left[x^{l-1+\frac{\a}{2}} f(x)\right]\Big|_\e^{+\infty} =0\,.
\ee
At $x=+\infty$, this expression vanishes because $f$ is assumed to be $L^2(\mathbb{R}_+,\vr_\a)$, whereas  at the origin it vanishes provided $f(x)$ vanishes at the origin with a power 
bigger than $1-\a/2$.  

In the unique case $l=1/2$, the Laplacian $\cK_2$ becomes the square of a differential operator:
\be
\cK_{\a,l}=(\cD^{\a,l})^2\qquad \Leftrightarrow\qquad l=\frac12\,.
\ee
In other words, the functions \Eq{cakx12} diagonalize the quadratic form \Eq{lk2} with differential operator $\cD_\mu:=\cD^{\a,1/2}_\mu$. In $D$ embedding dimensions,
\be
\boxd{\cD_\mu = \frac{1}{\sqrt{v_\a(x)}}\, \p_\mu \left[\sqrt{v_\a(x)}\,\,\cdot\,\right]\,,\qquad \cK_2=\cD_\mu\cD^\mu\,,}\label{DD}
\ee
\bs\ba
\cD_\mu c_\a(k,x)&=&-k_\mu\, s_\a(k,x)\,,\\
\cD_\mu s_\a(k,x)&=&k_\mu\, c_\a(k,x)\,.
\ea\es
Among the derivatives introduced, $l=1/2$ is the only case corresponding to a self-adjoint operator (with a suitable domain).

\subsubsection{Multi-fractional transforms?}\label{muft}

To complete the discussion, we would like to generalize to the multi-fractal space $\cE_*^D$. Before attempting that,
we make a remark about the generality of the results of sections \ref{frabe} and \ref{laqua}. In one dimension, they
are actually valid for any Lebesgue--Stieltjes measure weight $v_\a(x)\to v(x)$ such that
\be
|v(x)|=v(x)\,,\qquad D=1\,,
\ee
as a direct inspection of the invertibility of equation \Eq{fblt}, via \Eq{cakx}, shows. The requirement of positive definiteness is rather general and can include very irregular measures and measure weights of the form
\be\label{addi}
v(x)=\sum_\a g_\a v_\a(x)\,, \qquad g_\a,v_\a\geq 0\,.
\ee
However, in many dimensions we also require a much more restrictive property, namely, that the measure factorizes in the coordinates:
\be\label{facpro}
v(x)=\prod_\mu v_{(\mu)}(x^\mu)\,,
\ee
where the weights $v_{(\mu)}$ may differ from one another. This condition is fulfilled by real-order fractional measures, which are simple power laws, but it is not by weights of the type \Eq{addi}. Therefore, in $D\neq 1$ we do not expect to find an invertible transform on $\cE_*^D$ unless in very special cases, if any.

We can see this also in the alternative case where the sum over $\a$ is performed in front of the transform integral rather than in the measure and in the kernel functions $c_\a^l$ separately. In fact, since the multi-fractional measure \Eq{mume} is linear in $\vr_\a$ we can assume the multi-fractional Bessel transform to take the form
\be
\tilde f(k) := \sum_\a g_\a\int_0^{+\infty}\rmd\vr_\a(x)\,
f(x)\,c_\a^l(k,x)=: F_*^l[f(x)]\,,\label{mfst1}
\ee
with inverse
\be
f(x) = \sum_\a \tilde g_\a\int_0^{+\infty}\rmd\vr_\a(k)\,\tilde
f(k)\,c_\a^l(k,x)\,,\label{mfstau}
\ee
where $[\tilde g_\a]=-[g_\a]=-D\a$. As before, one plugs equation \Eq{mfst1} into \Eq{mfstau}:
\ba
\label{multifracid}
f(x) &=& \sum_{\a,\a'} g_\a\tilde
g_{\a'}\int_0^{+\infty}\rmd\vr_\a(k)\,c_\a^l(k,x)\int_0^{+\infty}\rmd\vr_{\a'}
(x')\, f(x')\,c_{\a'}^{l'}(k,x')\nonumber\\
     &=& \sum_{\a,\a'} g_\a\tilde
g_{\a'}\int_0^{+\infty}\rmd\vr_{\a'}(x')\,\cI_{\a,\a'}(x,x')\,f(x'),\label{mff}
\ea
where the multiple integral
\be
\cI_{\a,\a'}(x,x') := \int_0^{+\infty}\rmd\vr_\a(k)\,c_\a^l(k,x)\,c_{\a'}^{l'}(k,x')
\ee
entails integrals of the form
\be\nonumber
x\int_0^{+\infty}\rmd
k\,k^{\frac{\a-\a'}{2}+1}J_{l}(kx)J_{l'}(kx')\,.
\ee
Notice that $l'$ may differ from $l$ since the order of the Bessel function can depend on $\a$. 
To get agreement with the left-hand side of equation \Eq{multifracid}, $\cI_{\a,\a'}(x,x')$ should be proportional to $\de_\a(x,x')$. For $\a=\a'$ ($l=l'$), indeed $\cI_{\a,\a}(x,x')=\de_\a(x,x')$. However, for $\a\neq \a'$ this integral exists (if ${\rm Re}[l+l'+(\a-\a')/2+1]>0$, which we assume) and is not equal to a fractional delta. This result would be in conflict with equation \Eq{multifracid}, unless  $\cI_{\a,\a'}(x,x')$ identically vanishes for $\a\ne \a'$. This can happen only for the specific choice of the parameters \cite[formul\ae\ 6.574.1--3]{GR}
\be\label{ll'}
l-l'-\frac{\a-\a'}{2}=-2n\qquad {\rm or}\qquad l'-l-\frac{\a-\a'}{2}=-2n\,,
\ee
where $n$ is a non-negative integer. If this is the case, we finally obtain
\be
\cI_{\a,\a'}(x,x')=\de_{\a,\a'}\,\de_\a(x,x')\,,
\ee
where the first is a Kronecker delta and \Eq{mff} is an identity provided the following condition holds:
\be\label{kas2}
\sum_\a g_\a\tilde g_\a=1\,.
\ee
Given some energy cut-off $E$, the natural interpretation of the dimensionless coupling constants
\be
\g_\a= g_\a E^{-D\a}\,,\qquad \tilde\g_\a= \tilde g_\a E^{D\a}\,,
\ee
is that of probability weights, such that
\be\label{kas1}
\sum_\a \g_\a=1\,,\qquad \sum_\a \tilde\g_\a=1\,.
\ee
If, moreover, $\g_\a=\tilde \g_\a$, equations \Eq{kas2} and \Eq{kas1} become
\be\label{kaco}
\sum_\a \g_\a^2=1\,,\qquad \sum_\a \g_\a=1\,.
\ee
If $\a$ takes continuum values, these expressions hold only if $\g_\a=1$. If $\a$ takes discrete values, equation \Eq{kaco} becomes the set of Kasner conditions, which implies (taking the square of the second and using the first)
\be
\sum_{\a<\a'}\g_\a\g_{\a'}=0\,.
\ee
Therefore, at least one $\g_\a$ must have opposite sign with respect to the others.

Checking the condition \Eq{ll'}, one sees that it is true only if $n=0$ and $l=\pm\a/2+q$, where $q\in\mathbb{C}$. In particular, the case corresponding to the Laplacian \Eq{k1c} admits a multi-fractional transform, while the special case $l=\pm1/2$ (and any other where $l$ does not depend on $\a$) does not. As a consequence, there does not exist a multi-fractional Bessel transform in spaces equipped with the Laplacian $\cK_2$, which is the only one of the family \Eq{ckal} that can be written as the square of a first-order differential operator. 


Before moving on, we stress that the \emph{Ansatz}
\be
v(x)=\prod_\mu\left[\sum_\a g_\a^{(\mu)} v_\a(x^\mu)\right]
\ee
would lead to an invertible transform but it is not clear whether this corresponds to a natural multi-fractal model. 
The reason is that, in this case, the dimensionality along each direction flows independently from the others, while we would expect that a given $D$-dimensional configuration be evolved as a whole throughout the probed scales. In other words, a multi-fractal in a given $D$-dimensional embedding should be realized by taking ``snapshots'' of the whole object at different scales rather than taking the product of $D$ multi-fractals in one-dimensional embeddings. In one-dimensional systems the two procedures collapse one into the other, and there exist invertible momentum transforms for any weight $v_\a$ and a suitable function space.

\subsubsection{Complex fractional transforms?}\label{coft}

Another important extension is to complex fractional models. Fractional calculus can be extended to complex orders, by replacing the real-valued order $\a$ in integro-differential operators with a complex power \cite{Kob41,Lov71,LMNN,OLMN,NLM,HLA}. In our case, the only change is the replacement of the measure weight \Eq{frames2} (or \Eq{memu}, with $x$ replaced by its absolute value $|x|$) by \cite{fra4,frc2}
\be\label{kap2}
\tilde v_\a(x)=\sum_{\om=-\infty}^{+\infty} C_\om v_{\a+\rmi\om}(x):=\sum_{\om=-\infty}^{+\infty} C_\om \frac{x^{\a+\rmi\om-1}}{\Gamma(\a+\rmi\om)}\,,
\ee
where $\a,\om\in\mathbb{R}$ and $C_\om$ are complex coefficients. There are two major reasons why to be interested in such a generalization. The first is mathematical: genuine fractals have complex geometry and harmonic structures, reflected in the oscillatory behaviour of their spectral function \cite{DIL,KiL,Tep05,Akk1,ABS,Kaj10,LvF}. These structures are reproduced or approximated by complex fractional measures \cite{LMNN,NLM}. The second reason is physical. Consider a model with just one pair of conjugate frequencies $\pm\om_*$ and $C_0=1$:
\be\nonumber
\tilde v_{\a}(x) =v_{\a,\om_*}(x):= \frac{x^{\a-1}}{\Gamma(\a)}+C\frac{x^{\a+\rmi\om_*-1}}{\Gamma(\a+\rmi\om_*)}+C\frac{x^{\a-\rmi \om_*-1}}{\Gamma(\a-\rmi\om_*)}\,,
\ee
where $C$ is real. This measure is real, since it can be recast as \cite{frc2}
\be
\tilde v_{\a,\om_*}(x) = x^{\a-1}\left[\frac{1}{\Gamma(\a)}+a_{\a,\om_*}\cos\left(\om_*\ln x\right)+b_{\a,\om_*}\sin\left(\om_*\ln x\right)\right]\,,\label{come}
\ee
where $a_{\a,\om_*}=2C{\rm Re}\left[1/\Gamma(\a+\rmi\om_*)\right]$ and $b_{\a,\om_*}=2C {\rm Im}\left[1/\Gamma(\a+\rmi\om_*)\right]$ are real coefficients. In order to make the arguments of the logarithms dimensionless, one should introduce a length scale $x\to x/\ell_\infty$, \cite{fra4,frc2} which we do not need to consider here. Fractional spacetimes with measure \Eq{come} display the phenomenon of logarithmic oscillations, appearing in many chaotic systems \cite{Sor98}. The log-period, in turn, is tightly associated with a discrete scale invariance (DSI) of the measure under the coordinate rescaling
\be\label{dsi}
x\,\to\, \la_{\om_*}^n x\,,\qquad  \la_{\om_*}:=\exp\left(\frac{2\pi}{\om_*}\right)\,,\qquad n\in\mathbb{Z}\,.
\ee
As a matter of fact, any fractional complex measure \Eq{kap2} where the frequencies are multiples of a given one,
\be\label{omst}
\om=m\om_*\,, \qquad m\in\mathbb{Z}\,,
\ee
possess a DSI up to a global rescaling. These types of fractional measures have a rich hierarchy of scales \cite{fra4,frc2}. Near the fundamental scale $\ell_\infty$, which can be identified with the Planck length \cite{ACOS}, the texture of spacetime is discrete, while at scales larger than the log-period one can take the average of the measure and the system acquires a set of continuous effective symmetries. This may open up the possibility to construct models of quantum gravity with a natural discrete-to-continuum transition.

After this brief introduction, we want to see if spaces endowed with the measure \Eq{kap2} admit a unitary momentum transform. The considerations of the previous section show that, in all special cases where $|\tilde v_\a|=\tilde v_\a$, a transform exists. For instance, taking $m=0,\pm1$, $C_0=\Gamma(\a)/2$ and $C_{\pm\om_*}=-\Gamma(\a\pm\rmi\om_*)/4$, one gets
\be\label{sin2}
\tilde v_\a(x)=\frac14\,x^{\a-1} (2-x^{\rmi\om_*}-x^{-\rmi\om_*})=x^{\a-1} \sin^2\left(\om_*\ln \sqrt{x}\right),
\ee
which is positive definite and has log-period $\pi/\om_*$. However, the same arguments suggest that the general answer is No, at least in the continuum, since equation \Eq{kap2} is not even real-valued. More specifically, as in the multi-fractional case integral cross terms do not give a delta distribution and prevent a continuous transform to be unitary. This does not mean that there exists no momentum transform in complex spaces; rather, if it exists it is not a naive generalization of the fractional Bessel transform. We can see this in a calculation in the continuum, which also hints at the intriguing possibility that both position and momentum spaces are, in fact, lattices.

Define
\be
\tilde f(k) := \sum_\om C_\om\int_0^{+\infty}\rmd\vr_{\a+\rmi\om}(x)\,
f(x)\,c_{\a+\rmi\om}^l(k,x)\,,\label{mfst1o}
\ee
which is the naive extension of the unilateral fractional Bessel transform. As candidate inverse, we choose not to take the complex conjugate of this expression (obtained by replacing $\om\to-\om$), since the sum over $\om$ is bilateral and the final conditions on the parameters will be unaffected. Also, one soon realizes that the real part of the complex exponent must be the same as in the position-space measure, otherwise one meets with the same obstructions as for the multi-fractional case (even diagonal integration terms would not give the delta):
\be
f(x) = \sum_\om \tilde C_\om\int_0^{+\infty}\rmd\vr_{\a+\rmi\om}(k)\,\tilde
f(k)\,c_{\a+\rmi\om}^l(k,x)\,,\label{mfstauo}
\ee
where the coefficients $\tilde C_\om$ may differ from the $C_\om$. As kernel functions, we use (consider $D=1$)
\be\label{cakxc}
c_{\a+\rmi\om}^{l}(k,x) :=\Gamma(\a+\rmi\om) (kx)^{1-\frac{\a+\rmi\om}{2}} J_{l}(kx)\,.
\ee
As before, we plug \Eq{mfst1o} into equation \Eq{mfstauo}:
\ba
f(x) &=& \sum_{\om,\om'} \tilde C_\om C_{\om'}\int_0^{+\infty}\rmd\vr_{\a+\rmi\om'}(x')\, f(x')\,\cI_{\om,\om'}(x,x')\,,\label{ooi}\\
\cI_{\om,\om'}(x,x') &=& \Gamma(\a+\rmi\om')\,x^{-\frac{\rmi\om}{2}}{x'}^{-\frac{\rmi\om'}{2}} (xx')^{\frac{1-\a}{2}} \left(\frac{x'}{x}\right)^{\frac12} I_{\om,\om'}(x,x')\,,\nonumber\\\\
I_{\om,\om'}(x,x')   &=& \int_0^{+\infty}\rmd k\, k^{\frac{\rmi(\om-\om')}{2}} (kx) J_{l}(kx) J_{l'}(kx')\,,\label{ii}
\ea
where we allowed $l'$ to be different from $l$ in case $l$ depends on $\om$. To get a decomposition of the unit (i.e., to get an invertible unitary transform) we can use the integral representation of the Dirac distribution in terms of Bessel functions when $\om=\om'$ ($l=l'$), but we should be able to make all non-diagonal terms vanish. This operation can be done provided the complex power of $k$ in equation \Eq{ii},
\be\label{copa}
k^{\frac{\rmi(\om-\om')}{2}}=\rme^{\frac{\rmi(\om-\om')}{2} {\rm Log} k}\,,
\ee
is reduced to a trivial phase. However, this is not possible and the anti-transform \Eq{mfstauo} is not the inverse of \Eq{mfst1o}.

Incidentally, notice that if $k$ were discrete the phase \Eq{copa} could be rendered trivial. Suppose the frequencies are of the form $\om=m\om_*$; the simplest non-trivial example is $m=0,\pm 1$, such as in the measure \Eq{come}. Then, if $k$ were discrete the term \Eq{copa} would be identically equal to 1 if
\be
k=\exp\left(\frac{4\pi n}{\om_*}\right)\,,\qquad n\in\mathbb{Z}\,.
\ee
Setting $l=l'$ would remove any cross term, so one could conclude that $I_{\om,\om'}(x,x')$ $=\de(x-x')$ for any pair of $\om,\om'$. Proceeding further, one would find exactly the same lattice condition for $x$ and an algebraic condition on the coefficients $C_\om$ and $\tilde C_\om$. For non-real or non-positive-definite measures with $\om=m\om_*$, the sites of the lattices would always lie on the crests or nodes of the logarithmic oscillations, so that the actual measure would be $\tilde v_{\a}\propto x^{\a-1}$, up to some constant. This would not trivialize the theory to the real case because the newly found measure has discrete support. These results are only heuristic since they are based on the integral representation of the Dirac distribution, while the discrete nature of position and momentum space indicate that a sum representation is needed for self-consistency. We do not pursue this subject further here. 


\subsection{Bilateral world}\label{btbil}

Upon the replacement \Eq{unibi}, the notion of parity becomes meaningful. The basis functions \Eq{cakx} do not have, in general, definite parity, but the power of $|kx|$ in equation \Eq{cakx} compensates the measure weight $|x|^{\a-1}$ and cross terms of the form $c_\a^l(k,x) c_\a^{-l}(k,x')$ cancel out for suitable values of $l$, as it happens in the ordinary Fourier transform where integrals of $\cos(kx) \sin(kx')$ vanish by parity. Explicitly, the kernel functions are
\ba
c_\a^{l}(k,x) &:=& \prod_\mu c_{\a,\mu}^{l}(kx)\nonumber\\
&:=& \prod_\mu\Gamma(\a) |k^\mu x^\mu|^{\frac{1-\a}{2}}(k^\mu x^\mu)^{\frac{1}{2}} J_{l}(k^\mu x^\mu)\,.\label{cakx2}
\ea
Selecting the allowed values of $l$ will yield the desired representation of the Dirac distribution. Define
\be\label{ea0}
\bE_\a^l(k,x):=\frac1{2^D}\left[c_\a^{-l}(k,x)+A c_\a^{l}(k,x)\right]\,,\qquad A\in\mathbb{C}\,,
\ee
and the bilateral fractional transform as
\bs\be
\boxd{\tilde f(k) := \int_{-\infty}^{+\infty}\rmd\vr_\a(x)\,
f(x)\,{\bE_\a^l}^*(k,x)=:F_\a^l[f(x)]\,,}\label{fbit}
\ee
\be
\boxd{f(x) = \int_{-\infty}^{+\infty}\rmd\vr_\a(k)\,\tilde
f(k)\,\bE_\a^l(k,x)\,.}\label{fstaue}
\ee\es
The goal is to find some $A$ and $l$ such that equation \Eq{fstaue} is indeed the inverse of equation \Eq{fbit}. Plugging the former into the latter,
\ba
f(x) &=& \int_{-\infty}^{+\infty}\rmd\vr_\a(k)\,\tilde f(k)\,\bE_\a^l(k,x)\nonumber\\
 &=& \int_{-\infty}^{+\infty}\rmd\vr_\a(x')\,
f(x')\int_{-\infty}^{+\infty}\rmd\vr_\a(k)\,{\bE_\a^l}^*(k,x')\,\bE_\a^l(k,x)\nonumber\\
     &=& \int_{-\infty}^{+\infty}\rmd\vr_\a(x')\,f(x')\cI(x,x')\,.\nonumber
\ea
To get an identity, the integral in $k$ must yield a bilateral fractional delta. Using equation \Eq{ea0},
\ba
\cI(x,x') &=& \int_{-\infty}^{+\infty}\rmd\vr_\a(k)\,{\bE_\a^l}^*(k,x')\,\bE_\a^l(k,x)\nonumber\\
          &=& \frac1{4^D}\int_{-\infty}^{+\infty}\rmd\vr_\a(k)\,[c_\a^{-l}(k,x)c_\a^{-l}(k,x')+|A|^2 c_\a^{l}(k,x)c_\a^{l}(k,x')]\nonumber\\
          &&+\frac1{4^D}\int_{-\infty}^{+\infty}\rmd\vr_\a(k)\,[A^*c_\a^{-l}(k,x)c_\a^{l}(k,x')+A c_\a^{l}(k,x)c_\a^{-l}(k,x')]\,.\nonumber\\\label{tempo}
\ea
Writing $J_l(z)=z^l\cJ_l(z)$ as in equation \Eq{Jnu}, the function $\cJ_l$ is even under a reflection $z\to -z$. The last line features integrals of the form
\be\nonumber
\int_{-\infty}^{+\infty}\rmd k\,k\,\cJ_l(kx)\cJ_{-l}(kx')=0\,,
\ee
where 
we omitted a factor dependent on $x$ and $x'$; all these contributions vanish because the integrands are odd.
The remainder of equation \Eq{tempo} is split into four terms. In $D=1$,
\ba
\cI(x,x') &=& \frac14\int_{-\infty}^{+\infty}\rmd\vr_\a(k)\,[c_\a^{-l}(k,x)c_\a^{-l}(k,x')+|A|^2 c_\a^{l}(k,x)c_\a^{l}(k,x')]\nonumber\\
          &=& \frac{1}{4}\frac{\Gamma(\a)}{|xx'|^{\frac{\a-1}{2}}}\int_{-\infty}^{+\infty}\rmd k\,\left[(xx')^{\frac{1}{2}-l} k^{1-2l}\cJ_{-l}(kx)\cJ_{-l}(kx')\right.\nonumber\\
          &&\qquad\qquad\qquad\left.+|A|^2 (xx')^{\frac{1}{2}+l}k^{1+2l}\cJ_l(kx)\cJ_l(kx')\right]\nonumber\\
          &=&\frac{1}{4}\frac{\Gamma(\a)}{|xx'|^{\frac{\a-1}{2}}}\int_0^{+\infty}\rmd k\,\left\{(xx')^{\frac{1}{2}- l} [1+\rme^{\rmi\pi(1-2l)}]k^{1-2l}\cJ_{-l}(kx)\cJ_{-l}(kx')\right.\nonumber\\
          &&\qquad\qquad\qquad\left.+|A|^2 (xx')^{\frac{1}{2}+l}[1+\rme^{\rmi\pi(1+2l)}]k^{1+2l}\cJ_l(kx)\cJ_l(kx')\right\}\nonumber\\
         &=&\frac{1}{4}\frac{\Gamma(\a)}{|xx'|^{\frac{\a-1}{2}}} \left(\frac{x'}{x}\right)^{\frac12}\int_0^{+\infty}\rmd k\,(kx)\left\{ [1+\rme^{\rmi\pi(1-2l)}]J_{-l}(kx)J_{-l}(kx')\right.\nonumber\\
          &&\qquad\qquad\qquad\left.+|A|^2[1+\rme^{\rmi\pi(1+2l)}] J_l(kx)J_l(kx')\right\}\nonumber\\
        &{=}& \cA \left(\frac{x'}{x}\right)^{\frac12}\frac{\Gamma(\a)}{|xx'|^{\frac{\a-1}{2}}}\,\de(x-x')\nonumber\\
          & {=} &\cA\frac{1}{v_\a(x)}\,\de(x-x')\,,\nonumber
\ea
where
\be
\cA:=\frac{1+\rme^{\rmi\pi(1+2l)}+|A|^2[1+\rme^{\rmi\pi(1-2l)}]}{4}\,.
\ee
Setting $|A|=1$, one has $\cA=[\sin(\pi l)]^2$, which is equal to 1 if, and only if, $l=n-1/2$, where $n$ is a natural number (remember that $l>-1$). If $A=\pm 1$, for these values of $l$ equation \Eq{ea0} would be real-valued and one would recover the unilateral case. We set instead $A=\rmi$. Therefore, the fractional bilateral transform \Eq{fbit} is invertible with inverse \Eq{fstaue} if
\bs\be
\boxd{\bE_\a^l(k,x):=\frac{1}{2^D}\left[c_\a^{-l}(k,x)+\rmi\, c_\a^{l}(k,x)\right]\,,}
\ee
\be
l=n-\frac12\,,\qquad n\in\mathbb{N}\,.
\ee\es
These functions are orthonormal with respect to the fractional measure \Eq{memu} and yield the bilateral representation of the fractional delta distribution:
\be\label{deaee}
\boxd{\int_{-\infty}^{+\infty}\rmd\vr_\a(k)\,\bE_\a^l(k,x)\,{\bE_\a^l}^*(k,x')=\de_\a(x,x')\,.}
\ee
In turn, equation \Eq{deaee} implies the validity of the Parseval identity:
\ba
\|\tilde f\|^2 &=& \int \rmd\vr_\a(k) \tilde f^*(k) \tilde f(k)\nonumber\\
               &=&  \int \rmd\vr_\a(x)\int \rmd\vr_\a(x') f^*(x)f(x')\int \rmd\vr_\a(k)\,\bE_\a^l(k,x)\,{\bE_\a^l}^*(k,x')\nonumber\\
                &=& \int \rmd\vr_\a(x) f^*(x)f(x)\nonumber\\
                &=& \|f\|^2\,,
                \ea
and the transformations are unitary.

When $n=1$, we are in the special case $l=1/2$, which we write without index $l$:
\bs\label{fbit12}\be
\boxd{\tilde f(k) := \int_{-\infty}^{+\infty}\rmd\vr_\a(x)\,
f(x)\,\bE_\a^*(k,x)=:F_\a[f(x)]\,,}\label{f1}
\ee
\be
\boxd{f(x) = \int_{-\infty}^{+\infty}\rmd\vr_\a(k)\,\tilde
f(k)\,\bE_\a(k,x)\,,}\label{f2}
\ee
\ba
\bE_\a(k,x)&:=&\frac12\left[c_\a(k,x)+\rmi s_\a(k,x)\right]\nonumber\\
&=&\frac{1}{\sqrt{v_\a(k)v_\a(x)}}\,\frac{\rme^{\rmi k\cdot x}}{(2\pi)^{\frac{D}{2}}}\,.\label{ea12}
\ea
\es
When $\a=1$, the fractional transform reduces to the Fourier transform \Eq{ft}, $\bE_1(k,x)=\rme^{\rmi k\cdot x}/\sqrt{2\pi}$.
The discussion on the family of Laplacian operators remains unaltered provided one adopts the definitions \Eq{ckasv} and \Eq{quaf0}, where $v_\a$ is given by equation \Eq{memu}. In this way, factors with absolute value cancel appropriately.

All the results of \cite{fra4,frc1,frc2,ACOS} can be extended straightforwardly to a bilateral
world; in particular, the spectral dimension is the same, as already stressed on the grounds that $\ds$ is determined by the order of the Laplacian, not by the type of momentum transform employed.

Finally, the generalization to a multi-fractional measure follows exactly the same steps of the unilateral case, and fails for the same reason: integrals of the form
\be\nonumber
\int_{-\infty}^{+\infty}\rmd k\,|k|^{\frac{\a-\a'}{2}+1}J_{l}(kx)J_{l}(kx')
\ee
do not yield a delta for $\a\neq\a'$. Notice that $l$ is $\a$-independent, so there are no exceptions to this conclusion. The same obstruction occurs with complex measures.


\section{Conclusions}\label{conc}

In this paper, we have defined the momentum space dual to fractional spaces (a realization of fractal geometry being developed for applications to quantum gravity \cite{fra4,fra1}) via an infinite family of unitary bijections generalizing the Fourier transform. The kernel functions of these transforms are eigenfunctions of a family of Laplacian operators, only one of which is the square of a self-adjoint operator. This opens up the possibility to apply the tools of spectral analysis both to quantum mechanics and to quantum field theories living on fractional spacetimes \cite{frc2}. As a first direct application, in a companion paper we show the existence of well-defined quantum mechanics on such spaces, proving Heisenberg's principle and considering the standard example of the harmonic oscillator \cite{CNS}. The case of complex fractional measures, which typically display a discrete scale invariance, with require further study.

We now return to an assumption made in Section \ref{gese}, namely, the uniquess of the resolution of the identity. In other words, we required that the fractional Dirac distribution be the same in position and momentum space. In turn, this is tantamount to allowing the momentum transform to be an automorphism. If, however, the momentum measure $\tau(k)\neq\vr(k)$, the momentum transform maps different spaces one onto the other. The resulting fractional transform is still unitary and invertible, and all the above formul\ae\ hold upon replacing everywhere
\be
v_\a(k)\to v_{\bar\a}(k)\,,
\ee
where $\bar\a$ can differ from $\a$. Then, the kernel functions are no longer symmetric in $x$ and $k$. Even more generally, the key condition to hold is equation \Eq{conj} both in position and momentum space, so that one can work with arbitrary Lebesgue--Stieltjes measure weights $v(x)$ and $w(k)$ such that they are positive definite and the coordinate/momentum dependence factorizes along the $D$ directions. For instance, the transform \Eq{f1}--\Eq{f2} and the weighted plane waves \Eq{ea12} of the bilateral world become
\bs\ba
\tilde f(k) &:=& \int_{-\infty}^{+\infty}\rmd^Dx\,v(x)\,f(x)\,K^*(k,x)\,,\\
f(x)   &=& \int_{-\infty}^{+\infty}\rmd^Dk\,w(k)\,\tilde f(k)\,K(k,x)\,,\\
K(k,x) &=& \frac{1}{\sqrt{w(k)v(k)}}\,\frac{\rme^{\rmi k\cdot x}}{(2\pi)^{\frac{D}{2}}}\,,
\ea\es
while the fractal Dirac distributions in position and momentum space are
\be
\de_v(x,x')=\frac{\de(x-x')}{\sqrt{v(x)v(x')}}\,,\qquad \de_w(k,k')=\frac{\de(k-k')}{\sqrt{w(k)w(k')}}\,.
\ee
The Parseval relation follows through.

In the fractional case, the Hausdorff dimension of a momentum space with weight $v_{\bar\a}(k)$ is $\dh(\cE^D_k)=D\bar\a=(\bar\a/\a)\dh(\cE^D)$. On the other hand, the spectral dimension heavily depends on the details of the diffusion equation. The construction of a family of momentum spaces and transforms will be an important tool to solve the diffusion equation and, hence, to verify equation \Eq{conj} via an explicit construction.


\end{document}